\documentclass[fleqn,usenatbib]{mnras}

\usepackage{newtxtext,newtxmath}
\usepackage[T1]{fontenc}
\usepackage{ae,aecompl}

\usepackage{natbib}
\usepackage{bm}

\usepackage{graphicx}	
\usepackage{amsmath}	
\usepackage{amssymb}	

\usepackage{hyperref}

\title[ML Cluster Masses]{An application of machine learning techniques to galaxy cluster mass estimation using the MACSIS simulations}

\author[T. J. Armitage et al.]{Thomas J. Armitage,$^{1}$\thanks{E-mail: thomas.armitage-3@postgrad.manchester.ac.uk}
Scott T. Kay$^{1}$
and David J. Barnes$^{2}$
\\
$^{1}$Jodrell Bank Centre for Astrophysics, School of Physics and Astronomy, The University of Manchester, Manchester M13 9PL, UK\\
$^{2}$Department of Physics, Kavli Institute for Astrophysics and Space Research, Massachusetts Institute of Technology, Cambridge, MA 02139, USA\\
}

\date{Accepted XXX. Received YYY; in original form ZZZ}

\pubyear{2018}

\begin{document}
\label{firstpage}
\pagerange{\pageref{firstpage}--\pageref{lastpage}}
\maketitle

\begin{abstract}
Machine learning (ML) techniques, in particular supervised regression algorithms, are a promising new way to use multiple observables to predict a cluster's mass or other key features. To investigate this approach we use the \textsc{MACSIS} sample of simulated hydrodynamical galaxy clusters to train a variety of ML models, mimicking different datasets. We find that compared to predicting the cluster mass from the $\sigma -M$ relation, the scatter in the predicted-to-true mass ratio is reduced by a factor of 4, from $0.130\pm0.004$ dex (${\simeq} 35$ per cent) to $0.031 \pm 0.001$ dex (${\simeq} 7$ per cent) when using the same, interloper contaminated, spectroscopic galaxy sample. Interestingly, omitting line-of-sight galaxy velocities from the training set has no effect on the scatter when the galaxies are taken from within $r_{200c}$. We also train ML models to reproduce estimated masses derived from mock X-ray and weak lensing analyses. While the weak lensing masses can be recovered with a similar scatter to that when training on the true mass, the hydrostatic mass suffers from significantly higher scatter of ${\simeq} 0.13$ dex (${\simeq} 35$ per cent). Training models using dark matter only simulations does not significantly increase the scatter in predicted cluster mass compared to training on simulated clusters with hydrodynamics. In summary, we find ML techniques to offer a powerful method to predict masses for large samples of clusters, a vital requirement for cosmological analysis with future surveys.
\end{abstract}




\section{Introduction}
Galaxy clusters are the largest gravitationally bound structures in the Universe, tracing the high mass tail of the halo mass function, which makes them a valuable cosmological probe (see \citealt{Allen2011,Kravtsov2012,Weinberg2013,Mantz2014}). A key requirement for clusters to be used as a cosmological probe is a robust way to determine their masses. With impending surveys from the likes of eBOSS, DESI, eROSITA, SPT-3G, ActPol, and \textit{Euclid}, the number of clusters with high quality data is set to increase substantially, creating the need for an automated way to estimate their masses. To date, significant work has been put into developing physically motivated methods to determine the mass of a cluster. These methods can be broadly split into three classes: X-ray, gravitational lensing, and dynamical analysis of cluster galaxies.

By using data from X-ray observations of the diffuse intra-cluster medium (ICM) and assuming hydrostatic equilibrium, one is able to recover a measure of the cluster mass. However, sources of non-thermal pressure from active galactic nuclei (AGN), substructure, turbulence, magnetic fields, and cosmic rays introduce bias, estimated to be 10-40 per cent (\citealt{Lau2009,Rasia2012,Nelson2014,Henson2017}). A weak lensing analysis is likewise affected by cluster triaxilality and projection effects, but with a smaller bias of ${\sim}5$ per cent \citep{Okabe2010,Mahdavi2013,Hoekstra2015,Kettula2015,Oguri2011,Becker2011,Bahe2012,Henson2017}. Dynamical mass estimates use galaxies as tracers of the underlying dark matter (DM) distribution. To what extent galaxies are fair tracers of the velocity distribution is disagreed upon, with the galaxy velocity bias being within $\pm 10$ per cent of the DM dispersion (e.g. \citealt{Munari2013,Elahi2017,Armitage2018}). \cite{Old2013} and \cite{Old2014} performed a comparison of different dynamical mass estimators, finding that all dynamical methods suffer from significant scatter, typically ${\sim}0.3$ dex (i.e. a factor of two) when including projection effects and interloper galaxies.

Clusters are fundamentally complex systems, and many methods require strong assumptions to be made about the state of the cluster, such as spherical symmetry and virial or hydrostatic equilibrium. In addition, much of the information in the data is unused; for instance, in the case of the velocity dispersion - mass relation, the only information used is the tracer velocity dispersion. One would ideally build a model using all available information from the velocity, position and brightness distributions of the tracer population. Machine learning (ML) techniques offer a way of easily incorporating a set of cluster properties, referred to as \textit{features} in ML, into a model which once trained, can then be used to predict the masses of clusters. The caveat is that one abandons any physical model of the cluster. The first application of ML techniques to the problem of cluster mass estimation was performed by \cite{Ntampaka2015} and \cite{Ntampaka2016}, using the dark matter only (DMO) Multidark simulation. \cite{Ntampaka2015} use a support distribution machine (SDM, \citealt{Sutherland2012}) to characterise four feature distributions. Comparing to the velocity dispersion - mass relation, \cite{Ntampaka2015} found the ML implementation improves the scatter in the predicted mass by a factor of two. When the cluster galaxy samples are contaminated with interloper galaxies, the ML method still performed strongly, with less scatter in the contaminated sample for the ML model than for the uncontaminated $\sigma-M$ relation \citep{Ntampaka2016}. \cite{Ntampaka2018} has also explored the use of Convolutional Neural Networks (CNNs), often used in image processing, to predict cluster masses from mock X-ray observations of clusters. They find that using CNNs, reduces the scatter in the predicted mass to 12 per cent, down from $15-18$ per cent using traditional X-ray luminosity based methods.

In this work, we incorporate the impact of astrophysical processes by using the hydrodynamically simulated \textsc{MACSIS} \citep{Barnes2017} sample of 390 galaxy clusters, to develop a pipeline to predict cluster masses from galaxy data. The aim of this work is to show how one could apply ML techniques in an observational setting. This includes incorporating projection effects and training the ML models on imperfect masses derived from both X-ray and weak lensing analysis as well as quantifying the reduction in scatter when including additional data, such as the X-ray temperature and Sunyaev-Zel'dovich signal. We also test the impact of hydrodynamics by training ML models on both DMO and hydrodynamical simulations. The paper is organised as follows. Section \ref{sims} details the \textsc{MACSIS} clusters used in this analysis, while Section \ref{tech} outlines the analysis pipeline, including feature extraction. We compare several different ML algorithms in Section \ref{MLPerformance} to see which recover the mass of the clusters with the least scatter. In Section \ref{res} we present our main results, comparing different cluster feature sets and target masses. We also discuss the systematics introduced when training using DMO simulations and applying the model to hydrodynamical simulations. Finally we summarise and discuss our findings, explicitly stating the caveats of our work, in Section \ref{conc}.

\section{Simulation datasets} \label{sims}

The \textsc{MACSIS} sample consists of 390 galaxy clusters, with a mass range of $10^{14} < M_{200c} $\footnote{We define $M_{\rm 200c}$ as the mass enclosed within a sphere of radius $r_{\rm 200c}$ whose mean density is $200$ times the critical density of the Universe.}$ /\mathrm{M_\odot} < 4.75 \times 10^{15}$ at $z=0$. The sample is described in detail in \cite{Barnes2017}; here, we provide a brief overview. The \textsc{MACSIS} clusters were selected from a large $3.2$ Gpc DMO parent simulation. Clusters with a friends-of-friends mass between $10^{15} < M_{\rm FoF}/\mathrm{M_\odot} <10^{16}$ at $z=0$ were placed into logarithmic mass bins of width $\Delta \log_{10} M_\mathrm{FoF} =0.2$. The three lowest mass bins contained over 100 clusters each and so each bin was further subdivided into 10 logarithmically spaced sub-bins and 10 clusters randomly selected from each. For $M_\mathrm{FoF} > 10^{15.6} \; \mathrm{M_\odot}$ there were only 90 clusters, all of which were selected for the \textsc{MACSIS} sample. 

The selected haloes were then re-simulated using the zoom technique \citep{Katz1993,Tormen1997}, where the cluster region is simulated at a higher resolution than the parent box, with a lower resolution region beyond which accounts for the tidal forces from the surrounding large scale-structure. The \textsc{MACSIS} clusters were simulated using the \textit{Planck} cosmology \citep{Planck2014I}: with $\Omega_\mathrm{m}=0.3070$, $\Omega_\mathrm{b}=0.04825$, $\Omega_\mathrm{\Lambda}=0.6914$, $\sigma_8=0.8288$, $n_\mathrm{s}=0.9611$ and $h=0.6777$. Two zoom simulations of each cluster were created: a DMO simulation, \textsc{MACSIS-DMO}, with a particle mass of $5.2 \times 10^9 \, \rm M_\odot$, and a hydrodynamical simulation, \textsc{MACSIS-GAS}. For the latter, the initial gas and dark matter particle masses were $8\times 10^8 h^{-1} \, \rm M_\odot$ and $4.4\times 10^9 h^{-1} \, \rm M_\odot$ respectively. The Plummer equivalent gravitational softening length was fixed to $4 h^{-1} \, \rm kpc$ in physical units between $0<z<3$ and $16 \, h^{-1} \, \rm kpc$ in comoving coordinates at higher redshift.

The code and sub-grid physics model used in the \textsc{MACSIS-GAS} sample is the same as the BAHAMAS simulation \citep{McCarthy2017}. The simulations were performed using the smoothed particle hydrodynamics code \textsc{gadget-3}, last publicly discussed in \cite{Springel2005}. \textsc{gadget-3} has been adapted to include the \textsc{OWLS} sub-grid physics model \citep{Schaye2010}, which incorporates the effects of radiative cooling \citep{WiersmaSchayeSmith2009}, star formation \citep{SchayeDallaVecchia}, and feedback from supernovae \citep{Dalla2008} and AGN \citep{BoothSchaye2009}. The feedback calibration for BAHAMAS was done to match the observed gas mass fraction of groups and clusters \citep{Vikhlinin2006,Maughan2008,Sun2009,Pratt2009,Lin2012} and the global stellar mass mass function \citep{Li2009,Baldry2012,Bernardi2013}, as described in \cite{McCarthy2017}. \cite{Barnes2017} further showed that the \textsc{MACSIS-GAS} clusters reproduce the observed mass dependence of the hot gas mass, X-ray luminosity and Sunyaev-Zel'dovich signal at redshift $z=0$. 

In this paper we define any self-bound object as identified by \textsc{subfind} \citep{Springel2001,Dolag2009} with a stellar mass greater than $10^{10} \, \rm M_\odot$ as a galaxy. \cite{Henson2017} found that due to the initial \textsc{MACSIS} cluster selection being based on $M_\mathrm{FoF}$, the lower mass clusters in the sample had systematically low concentrations and high spin parameters compared to the rest of the sample. To ensure that we do not use unrepresentative low-mass clusters, we perform the same mass cut as \cite{Henson2017}, removing clusters below $M_{200c} = 10^{14.5} h^{-1} \rm M_\odot$. This reduces the \textsc{MACSIS} sample to 357 clusters at redshift $z=0.25$, which is the sole redshift considered in our analysis.

\section{Machine learning methods} \label{tech}
Currently, scaling relations are widely used to cheaply obtain cluster mass estimates (e.g. \citealt{Arnaud2007,Evrard2008, Vikhlinin2009, Pratt2009,Zhang2011, Planck2014XX,Sereno2015,Mantz2016}). These scaling relationships must either be calibrated using simulations or by using a high quality dataset where more advanced, and data heavy, mass estimation techniques can be applied (e.g. using weak lensing data). Likewise, supervised machine learning techniques require a calibration (training) dataset where the true value of a variable (feature) is known. A key advantage of ML methods, however, is that they are easily generalised to use many different features together, potentially making better use of all the available information. 

Machine learning techniques can be broadly classified into two types: classification and regression. Classifiers aim to place objects into discrete groups, e.g. whether a galaxy cluster is relaxed or not. Regression algorithms are for continuous distributions, and so are more suited for cluster mass estimation, which is the main aim of this study. One of the simplest regression models is the familiar least-squares linear regression, which can be generalised to fit $n$ variables (features) to predict a desired observable. In this section we shall discuss how we extracted useful features from our cluster sample and the general principles of regression models before comparing the performance of different algorithms in Section \ref{MLPerformance}\footnote{We provide a basic template of the ML implementation used in this work, available at \url{https://github.com/TomArmitage/ML_Template}}.

\subsection{Feature extraction and selection}
\begin{figure}
	\centering
	\includegraphics[width=0.99\linewidth]{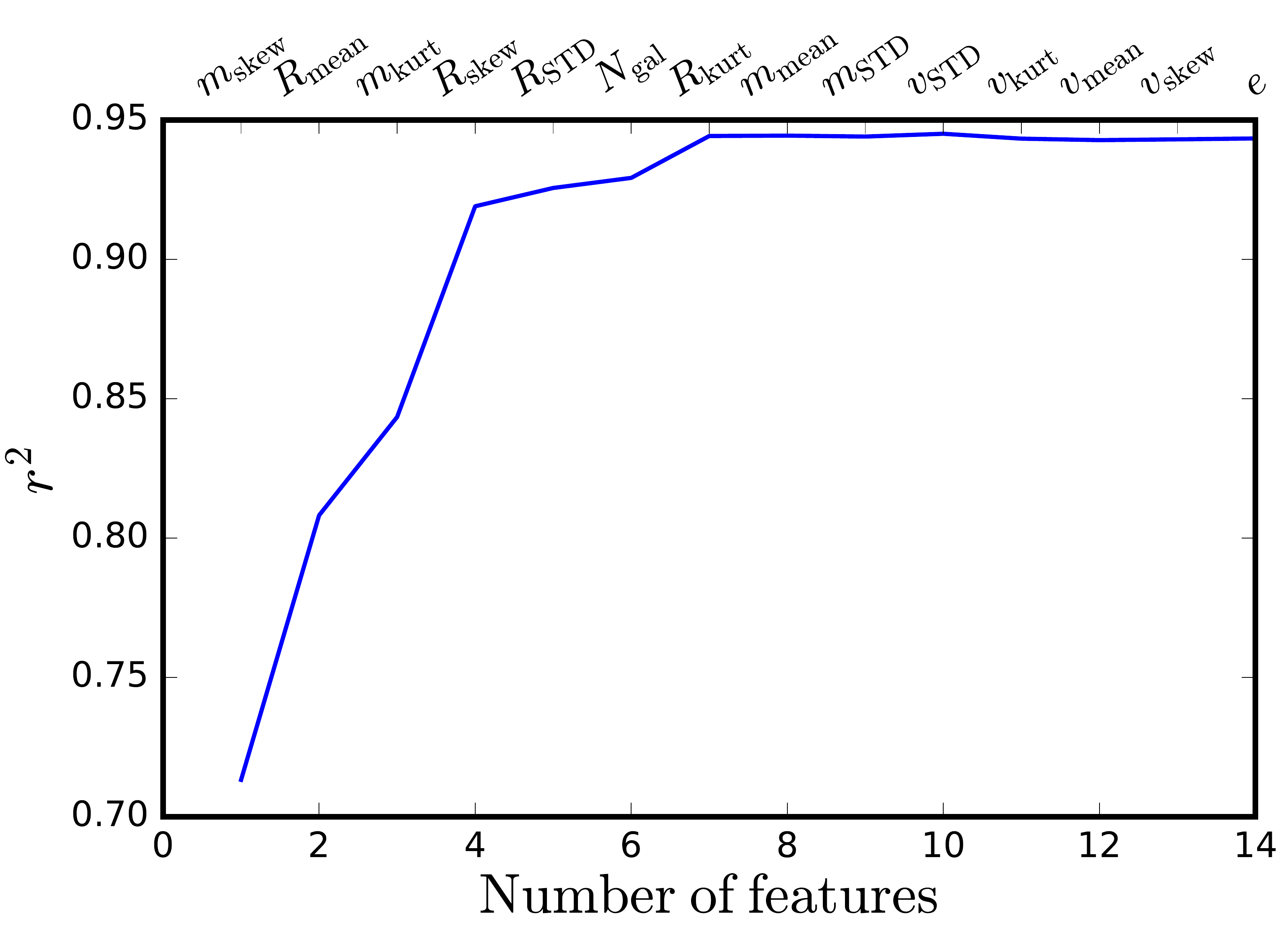}
	\caption{Coefficient of determination against number of features in the training sample. Initially all features are used to train a regression model, the feature contributing the least information is then discarded. The feature discarded is shown at the top of the plot (moving from right to left). The process is repeated until only one feature remains.}
	\label{fig:RFE}
\end{figure}

Before we can apply the regression models we must first extract features (observables) from the \textsc{MACSIS} clusters. A feature must be a scalar value for each cluster, such as the number of galaxies present inside a given aperture, or the cluster velocity dispersion. In this work we initially focus on features that can be obtained from an optical survey of clusters, i.e. projected galaxy positions, line-of-sight velocities and stellar masses. We project all cluster samples along a cylinder of length $10r_{200c}$ within a radius of either $r_{200c}$ or $1.5 \, \rm Mpc$, to contaminate the sample with interloper galaxies. The first aperture, by definition, scales with cluster mass ($r_{200c} \propto M_{200c}^{1/3}$), while the second is chosen to assess the effect of choosing a fixed physical aperture. For each of these cluster projections we calculate the mean, standard deviation, skewness and kurtosis of said galaxy properties. We also consider the total number of galaxies and the ellipticity (ratio of semi-minor to semi-major axes), $e$, using the method described in \cite{Bett2007}. The position information is encoded as the projected cluster-centric radius of each galaxy. The features chosen for this work are by no means the only ones possible; one could incorporate relaxation criteria, or morphological indicators for example, as well as information from other datasets, such as weak lensing shear or X-ray surface brightness maps. In this paper, we focus on relatively direct observables, with more complex features left for future work. For conciseness we define the PHOT and SPEC feature sets: PHOT contains only projected positional galaxy information and their respective stellar masses, while SPEC includes the PHOT features and line-of-sight velocity information. Later, we also incorporate the X-ray temperature, X-ray (bolometric) luminosity and SZ flux as additional features. We summarise the quantities in each feature set in Table \ref{tab:FeatBreakDown}.

\begin{table}
	\centering
	\caption{Summary of the quantities (features) in each feature set used throughout this paper. $R$, $m$ and $v$ are the projected radial distance, stellar mass and line-of-sight velocity of each galaxy in a cluster. We take the mean, standard deviation, skewness and kurtosis of these distributions as features for each cluster. The ratio $e$ of the semi-minor and semi-major axes is calculated following \protect\cite{Bett2007}. All features in X are the core-excised variants found in \protect\cite{Barnes2017}.}
	\label{tab:FeatBreakDown}
	\begin{tabular}{l l}
		\hline
		Feature set & Included Features \\
		\hline 
		PHOT: & $R_\mathrm{mean,\, std,\, skew,\, kurt}$, $m_\mathrm{mean,\, std,\, skew,\, kurt}$, $ N_\mathrm{gal}$, $e$\\
		SPEC: & PHOT + $v_\mathrm{mean,\, std,\, skew,\, kurt}$ \\
		X: & $T_\mathrm{500c}$, $L_\mathrm{X,500c}$\\
		SZ: & $Y_\mathrm{SZ,5r_{500c}}$\\
		\hline
	\end{tabular}
\end{table}

Having generated a number of features from the galaxy population of each cluster we must now determine which features contain the most information relevant to the target feature, in this case $M_{200c}$. Including features that are uncorrelated with the target output can act to wash out the information contained in other features, as well as increase the computation time unnecessarily. One way to select the best features is through Recursive Feature Elimination (RFE, \citealt{Guyon2002}). RFE is a brute force technique, that given a regression (or classifier) algorithm, trains using all features initially and then eliminates the least important feature according to some goodness of fit metric, proceeding recursively until a set number of features remain. The scoring metric used in this work is the coefficient of determination, $r^2$, defined as
\begin{equation}
r^2 (y,\hat{y})=1- \frac{\sum_{i=0}^{n_\mathrm{samples}-1} (y_i-\hat{y}_i)^2}{\sum_{i=0}^{n_\mathrm{samples}-1} (y_i-\bar{y})^2} \,
\end{equation}
where $\hat{y}_i$ is the predicted output for the $i^{th}$ cluster and $y$ is the set of true outputs, in this case cluster masses, with a mean value $\bar{y}$. $n_\mathrm{samples}$ is the total number of samples (galaxy clusters) used to train the ML model. The optimal $r^2$ score is 1, indicating a perfect fit to the data. Fig. \ref{fig:RFE} shows an example $r^2$ curve generated using the \textsc{MACSIS} sample, using galaxies within an aperture of $r_{200c}$ and the SPEC feature set. One can see that beyond 7 features the curve is very flat, indicating that those features are contributing little additional information, while the first few features greatly improve the goodness of fit. The maximum value of $r^2$ is obtained with 10 features with the inclusion of the velocity dispersion, and so in the pipeline we develop only those features that would be used to train the regression model (the features eliminated are $v_\mathrm{kurt}$, $v_\mathrm{mean}$, $v_\mathrm{skew}$ and $e$). Note that the only velocity feature present after RFE is the velocity dispersion, which we explore further in Section \ref{res}. However, RFE is limited in a number of ways. Namely the initial weights on the importance of different features can be unreliable due to the nature of the problem, so the exact order of the features is unstable, however the overall trends are meaningful. It also ignores any non-linear relationships between features. The $r^2$ profiles do not qualitatively change with different combinations of features, e.g. PHOT+X, SPEC+YZ; as may be expected, there is a lot of redundant information in our initial feature set.

Another way to reduce the number of features is to perform a principal component analysis (PCA). Put briefly, PCA takes the multi-dimensional feature space (where each galaxy cluster has a position set $[v_\mathrm{std}, m_\mathrm{skew}, N_\mathrm{gal},...]$) and re-projects it so that each new axis is independent of the others, i.e. the covariance matrix of the new features is diagonalised. One can then remove features based on their respective variance, allowing one to retain the maximum information after removing a set number of features. A significant disadvantage with PCA is that each new feature is a linear combination of all the original features, making it difficult to determine what type of data is most useful. In practice, we found little difference in the estimated mass when pre-processing features using either RFE or PCA. Due to additional abstraction with PCA we opt to use RFE, selecting the feature set with the highest $r^2$ score, throughout the rest of the paper unless explicitly stated otherwise. We discuss which features contain the most useful information in Section \ref{res}.

A second important preprocessing step is standardisation of the data, as many ML algorithms assume the features have approximately zero mean and a standard deviation of one. To ensure our data are robust to outliers we instead subtract the median of each feature and divide by the $1\sigma$ ($16^{\rm th}/ 84^{\rm th}$) percentile range. 
For properties whose values span orders of magnitudes we take the base 10 logarithm before standardising ($v_\mathrm{std}$, $N_\mathrm{gal}$, $Y_\mathrm{SZ,5r_{500c}}$, $T_\mathrm{500c}$ and $L_\mathrm{X,500c}$).

\subsection{Model training and testing} \label{modelDescription}
The next stage of the analysis is to use the feature set, after RFE, to train a model to predict some desired quantity or target feature, in this case cluster mass. To train a model we require both a training and testing sample. The target feature is known for the training sample and is used to train the model. The model is then used to predict the same target feature in the test sample. It is vitally important that the training sample is a fair representation of the underlying distribution. In cases with only a few hundred samples one would ideally like to obtain estimated masses for all clusters. This can be achieved using $k$-fold cross validation. $k$-fold cross validation randomly splits the available data into $k$ subsamples; $k-1$ subsamples are then used to train a model which is then tested on the remaining subsample. The subsamples are then rotated with all galaxy clusters tested exactly once. In this work we take the commonly used value of $k=10$, i.e. a 90-10 split \citep{Ball2010}. Once we have a predicted mass for each cluster we can then compute properties such as the scatter and bias between the predicted mass and the true mass. To prevent any duplication of data in the training and testing samples, we perform the $k$-fold cross validation completely separately for when we project clusters along the $x$, $y$ and $z$ axes. Note that there should be negligible bias using this method, as long as the training set is a fair sample of the underlying distribution (we checked this and found this to be a good assumption in practice). We therefore focus on comparing results for the logarithmic scatter, $\delta$, in $ \log_{10} \left( M_{\rm pred}/ M_{\rm true}\right)$, where $M_{\rm pred}$ is the predicted cluster mass from the ML model and $M_{\rm true}$ is the true mass of the cluster.


We now discuss the implementation of the ML algorithms used in this work. There are a wide variety of regression algorithms available through software packages such as the Python based \textsc{scikit-learn} used here \citep{sklearn}. We shall now describe two of the simpler, yet most relevant, procedures that are widely used throughout this work. For a brief overview of the other methods we considered and their performance, see Section \ref{MLPerformance}.

The simplest regression algorithm is ordinary linear regression (OLR), which is used ubiquitously. To summarise, if one has a set of $n$ independent observables of $i$ systems, $\mathbf{X}=\{\mathbf{x}_0^T,\mathbf{x}_1^T,...,\mathbf{x}_n^T\}$, where $\mathbf{x}$ is a vector of length $i$ with corresponding dependent variables $\mathbf{y}$, then one can relate the two with an unknown vector $\bm{\beta}$. By finding the value $\bm{\hat{\beta}}$ that minimises a loss function, in this case the sum of least squares
\begin{equation}
\mathcal{L}_\mathrm{RSS} = \sum_{i} (y_i - \mathbf{x}_i^{\rm T} \bm{\beta})^2 \; ,
\end{equation}
one tends to obtain a good estimate of the true value of $\bm{\beta}$. However, if the columns of $\mathbf{X}$ are highly correlated then the above procedure can yield highly unstable estimates of $\bm{\beta}$. For a simple example of how this can arise, if one has just two features, $\mathbf{x}_0$, and $\mathbf{x}_1=\mathbf{x}_0 + \mathbf{\delta}$, where $\mathbf{\delta}$ is a small random number, then there will be many values of $\bm{\hat{\beta}}$ which produce low values of RSS. This can result in over-fitting of the data, as the value of $\bm{\hat{\beta}}$ becomes very sensitive to the specific training values of $\mathbf{X}$.

A solution to this is to introduce a term which penalises large values of $\bm{\beta}$. In the case of ridge regression \citep{Hoerl1970} the quantity to minimise is
\begin{equation}
	\mathcal{L}_\mathrm{Ridge} = \mathcal{L}_\mathrm{RSS}(\bm{\beta}) + \lambda \sum_{j=1}^{n} \beta_j^2 \; ,
\end{equation}
where the second term prefers values of $\beta_j$ closer to zero. If one has already standardised the feature set by subtracting the mean and normalising by the standard deviation then this is good assumption. To find $\bm{\hat{\beta}}$ in the case of ridge regression one computes
\begin{equation}
\bm{\hat{\beta}} = \left( \mathbf{X}^T \mathbf{X} + \lambda \mathbf{I} \right)^{-1} \mathbf{X}^T \mathbf{y} \; ,
\end{equation}
where $\lambda$ is a hyper parameter that needs to be chosen beforehand. In the case of $\lambda=0$ then ridge regression becomes identical to OLR and as $\lambda \rightarrow \infty$, $\bm{\hat{\beta}} \rightarrow \mathbf{0}$. The selection of $\lambda$ is often made during the training process by testing a range of different values and comparing the residuals of $y-\hat{y}$. In this work we perform this search of $\lambda$ values via a basic grid search of the parameter space. During training, multiple values of $\lambda$ are tested on a separate subset of the training data to see which value of $\lambda$ minimizes the sum of residuals.

\section{Performance of ML algorithms} \label{MLPerformance}
\begin{figure}
	\centering
	\includegraphics[width=0.99\linewidth]{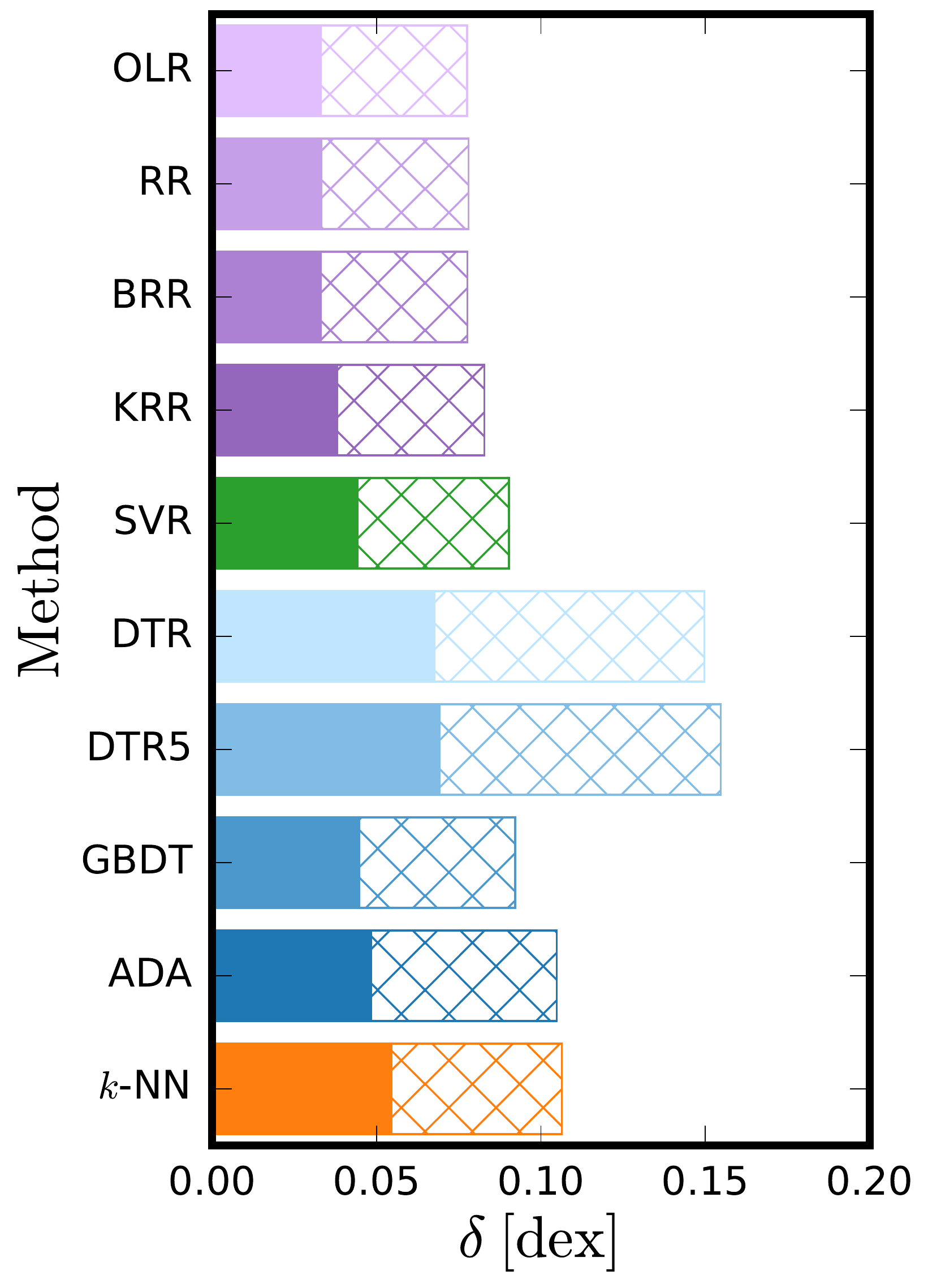}
	\caption{Comparison of the scatter between the predicted and true mass, $\delta$, for the ten different machine learning algorithms, using the SPEC feature set. The solid portion of each bar illustrates the $1\sigma$ scatter and the hatched section $2\sigma$. Similar methods are coloured as such, with acronyms defined in the text.}
	\label{fig:modelCompR200StelMass}
\end{figure}

In this section we compare a selection of regression algorithms to determine the most suitable algorithm for predicting the masses of our clusters. All methods were trained using the SPEC feature set to predict $M_{200c}$. If any cluster contained fewer than 10 galaxies we rejected it from the sample. Since we had already applied the mass cut of \cite{Henson2017} the latter condition only removed around one cluster for any given projection. We now provide a brief\footnote{For a more detailed description of each method and their implementation, see \cite{sklearn} and the documentation included with \textsc{scikit-learn}.} overview of each ML algorithm considered:

\begin{itemize}
	\item OLR - Ordinary Linear Regression. This method was already described in Section \ref{modelDescription}.
	\item RR, BRR, KRR - three variants of ridge regression, the basic implementation of which is also described in Section \ref{modelDescription}. The Bayesian ridge regression (BRR) variant incorporates Gaussian priors over the model weights. Kernel ridge regression (KRR) combines ridge regression with the kernel trick, used in SVR below.
	\item SVR - Support vector regression: The continuous form of support vector machines (SVMs), which classifies data by embedding the features into a higher dimensional space and separating different classes using a hyper-dimensional plane, which can also be thought of as re-projecting the data using some kernel function (this is referred to as the `kernel trick'). SVR outputs continuous values rather than discrete classes as for SVM.
	\item DTR - decision tree: A decision tree is created using a set of \textit{if-then-else} rules. The depth of the tree can be adjusted; for DTR5 we use a maximum depth of 5, which can help prevent the model from over-fitting the training data. We also allow the tree to go deep enough that all leaves only contain one sample (DTR).
	\item GBDT - gradient boosted regressor trees: Gradient boosting involves sequentially adding trees of increasing complexity in order to minimise some loss function. 
	\item ADA - AdaBoosted decision trees: AdaBoosted trees are an ensemble of small decision trees, which are weighted based upon their accuracy, with additional trees added to improve poor estimates. The final prediction is a weighted sum of all decision trees \citep{Freund1997}.
	\item $k$-NN - $k$ Nearest Neighbours: After computing the $k$ nearest neighbours to the desired point in feature space, the average target value of the $k$ neighbours is taken as the estimate.
\end{itemize}

For our main results, we present the scatter in the predicted-to-true mass ratio, $\delta$, which we define as half the $16^{\rm th} - 84^\mathrm{th}$ percentile range of
\begin{equation}
	\mathcal{R} = \log_{10} \left( M_\mathrm{pred}/ M_\mathrm{true} \right),	
\end{equation}
where $M_\mathrm{pred}$ is the predicted cluster mass and $M_\mathrm{true}$ is the mass we are attempting to recover, usually $M_{200c}$. In general, we find the distribution of $\mathcal{R}$ to be approximately Gaussian, and we show the $1\sigma$ (68 per cent) and $2\sigma$ (95 per cent) percentile scatter in Fig. \ref{fig:modelCompR200StelMass} for each method as the solid and hatched bars respectively. For each model, we predict $M_{200c}$ for each cluster via $k$-fold cross validation. The analysis was repeated for three orthogonal projections of the clusters and the predicted masses of the tested clusters (around 1,000 in total) were used to calculate the scatter, $\delta$ in dex. 

We can see from Fig. \ref{fig:modelCompR200StelMass} that the best performing models belong in the ridge regression family: RR, BRR, and KRR. However, KRR suffers from a reasonable number of catastrophically-underestimated outliers. These occur at the edges of the cluster mass distribution and are the primary cause of the slightly increased scatter seen in Fig. \ref{fig:modelCompR200StelMass}. Due to these outliers we discard KRR. Surprisingly, ordinary linear regression is also one of the best performing methods, even though it is the simplest algorithm tested.

All of these models have a $1 \sigma$ scatter of less than 0.075 dex with the best having $\delta {\sim}0.03$ dex. Compared to the results of \cite{Armitage2018Proj2}, who quantified the scatter of traditional dynamical mass estimators using the high resolution hydrodynamical C-EAGLE clusters \citep{Barnes2017b}, such as the caustic and Jeans methods, all of the machine learning models outperform the traditional estimators under comparable conditions which produce an average scatter of $0.1$ dex, i.e. a factor of 3 reduction. Furthermore, when using a stellar mass limited sample of $10^9 \, \mathrm{M_\odot}$, with perfect knowledge of 3D galaxy positions, velocities and cluster membership \cite{Armitage2018Proj2} finds the scatter is only reduced to 0.07 dex, significantly higher than the 0.03 dex obtained with the projected and contaminated \textsc{MACSIS} sample.

The OLR, RR and BRR methods produce the least scatter with little to distinguish them. Both the RR and BRR methods require hyper-parameters to be tuned, whereas OLR has none. RR has one hyper-parameter compared to four for BRR, which reduces the computational time required to optimise the parameters via a grid-search. We performed all of the analysis in Section \ref{res} using both RR and OLR, finding little difference between the two algorithms, with OLR suffering from slightly more scatter. As a result we only present results obtained using the ridge regression method (RR) from now on.


\section{Feature Set Comparison} \label{res}
Having selected ridge regression as the optimal algorithm to predict the masses of our clusters, we now consider the performance of different feature sets and the effect of baryonic physics. We first consider the performance of different optical feature sets and aperture sizes, namely the difference between a photometric and spectroscopic galaxy sample, when the galaxies are selected inside $r_{200c}$ or within a fixed aperture of $1.5 \, \rm Mpc$. We then add X-ray and SZ properties, before training the models to fit either estimated weak lensing or hydrostatic masses. Finally, we test the sensitivity of our ML model to baryonic physics by training on a DMO simulation and applying that model to hydrodynamic simulations.

\subsection{Dependence on optical features and aperture size}
\begin{figure}
	\centering
	\includegraphics[width=0.95\linewidth]{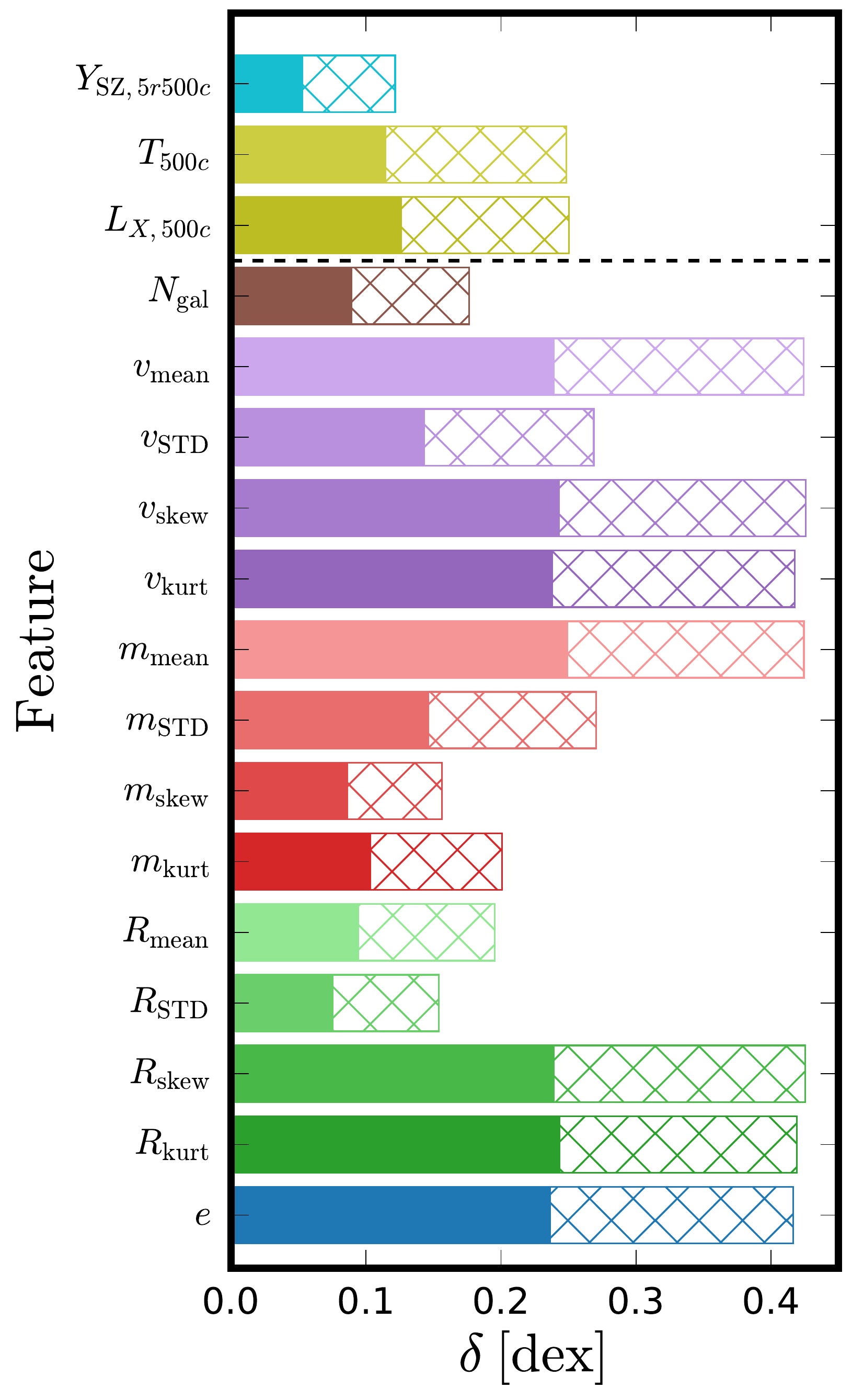}
	\caption{Comparison of the scatter, in dex, of each individual feature considered in this work. The scatter is defined as in Fig. \ref{fig:modelCompR200StelMass}. The cluster sample was selected within an aperture of $r_{200c}$. The different colours correspond to the physical feature; green, red and purple denote features derived from the galaxy radial, mass and velocity distribution of each cluster, respectively. The number of galaxies, ellipticity, X-ray features and the SZ flux are shown in brown, blue, yellow and cyan respectively. Features derived from the cluster ICM (galaxies) are above (below) the black dashed line.}
	\label{fig:FeatInd}
\end{figure}
\begin{figure}
	\centering
	\includegraphics[width=0.95\linewidth]{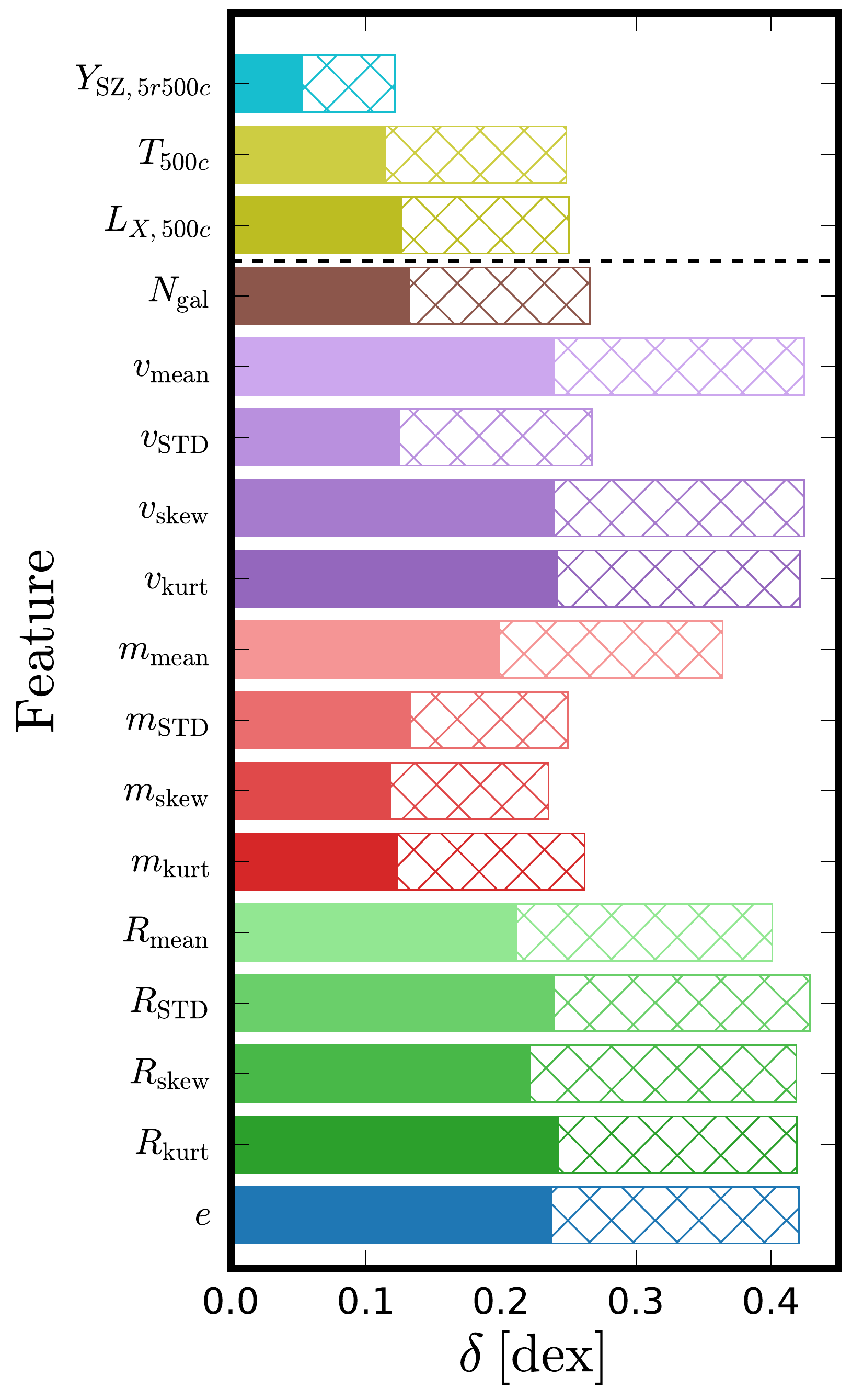}
	\caption{Comparison of the scatter, in dex, of each individual feature considered in this work measured within a fixed aperture of 1.5 Mpc for each cluster, instead of $r_{200c}$ as in Fig. \ref{fig:FeatInd}.}
	\label{fig:FeatInd_Phys}
\end{figure}

\begin{table}
	\centering
	\caption{The $1\sigma$ percentile scatter, $\delta$, in $\log_{10} \left( M_{\rm est}/M_{200c} \right)$ for different feature sets. Table \ref{tab:FeatBreakDown} shows the features included in each row. Galaxies were either taken within a cluster aperture of $r_{200c}$ or $1.5 \, \rm Mpc$. For comparison, we provide the scatter when obtaining masses via the $\sigma-M$ relation, also calibrated using the \textsc{MACSIS} clusters. The errors in $\delta$ are the $1 \sigma$ percentile scatter obtained via bootstrap resampling of the clusters.}
	\label{tab:PhotVSpec}
	\begin{tabular}{lcc}
		\hline
		$\mathrm{Data \, set}$ & $\delta_{r_{200c}}$   &   $\delta_{1.5 \, \rm Mpc}$ \\
		\hline
		PHOT                     & $0.032\pm0.001$          & $0.065\pm0.002$          \\
		SPEC                     & $0.031\pm0.001$          & $0.064\pm0.002$          \\
		$\sigma-M$                 & $0.130\pm0.004$          & $0.122\pm0.003$          \\
		\hline
	\end{tabular}
\end{table}

We first compare how constraining the different galaxy features are individually. We train an RR model to predict $M_{200c}$ from a single feature using $k$-fold cross validation with 10 bins. Fig. \ref{fig:FeatInd} shows the scatter in the resulting mass estimates, where the galaxies were selected from inside a projected aperture of $r_{200c}$. We find that $R_{\rm STD}$, $m_\mathrm{skew}$ and $N_\mathrm{gal}$, are the most constraining galaxy-based features while $e$, an indicator of cluster morphology, is one of the least constraining features with a scatter of $0.23$ dex, as the shape of a cluster is weakly correlated with mass, i.e. both small and large clusters are equally likely to be extended. We note that if we train on completely random noise we obtain a scatter of ${\sim} 0.24$ dex, as the model effectively outputs the mean mass of the training sample, so no feature should ever result in a scatter $>0.24$ dex. One caveat is that \textsc{MACSIS} does not include low mass clusters, which may be more likely to be dynamically relaxed and spherical. As might be expected, the velocity dispersion is the single most constraining velocity feature ($\delta \sim 0.143$ dex), however, it is significantly less constraining than $R_{\rm STD}$ ($0.075$ dex), $m_\mathrm{skew}$ ($0.086$ dex), $ N_\mathrm{gal}$ ($0.089$ dex), $R_\mathrm{mean}$ ($0.094$ dex) and $m_\mathrm{kurt}$ ($0.103$ dex). 

If instead we take galaxies within a fixed aperture of 1.5 Mpc, we obtain the results shown in Fig. \ref{fig:FeatInd_Phys}. Selecting a fixed physical aperture removes the intrinsic mass dependence that is included when using $r_{200c}$. The most marked change is the increased scatter for $R_{\rm mean}$ and $R_\mathrm{STD}$, by a factor of 2 and 3, respectively. Many of the other galaxy-derived features also suffer an increase in scatter, for similar reasons. However, both $v_\mathrm{STD}$ and $m_\mathrm{mean}$ improve with the fixed aperture, with their scatter decreasing to $\delta =0.87$ and $\delta =0.8$ of the $r_{200c}$ values, respectively. This is likely due to the fraction of interlopers present in the $1.5$ Mpc aperture being half that of the $r_{200c}$ sample on average (we note that the mean $r_{200c}$ of the \textsc{MACSIS} clusters is ${\sim} 1.9$ Mpc).

We now combine the features into two sets that represent different observational sets, PHOT and SPEC, using the two different aperture sizes (see Table \ref{tab:FeatBreakDown}). We remind the reader that the features were first filtered using RFE and then used to train ridge regression models using the true $M_{200c}$ cluster mass. Table \ref{tab:PhotVSpec} shows the scatter, $\delta$, for each feature set. We see that there is remarkably little difference between PHOT and SPEC results, although both are significantly better than using the $\sigma -M$ relation, which is included in Table \ref{tab:PhotVSpec} for comparison. (We note that we fit the $\sigma-M$ relation using the \textsc{MACSIS} galaxies, and then use that relation to predict the mass of \textsc{MACSIS} clusters from their velocity dispersion.)

Interestingly the fixed $1.5 \, \rm Mpc$ aperture results in an increased scatter (by ${\sim} 0.03$ dex) for the PHOT and SPEC set, but decreases the scatter by ${\sim}0.01$ dex for the $\sigma-M$ relation. The decrease in scatter for the $\sigma-M$ relation is far less statistically significant, at around the $2 \sigma$ level. The increase in scatter for the machine learning methods comes from the loss of information from galaxies beyond $1.5 \, \rm Mpc$, particularly for the more massive systems. In addition to this, $r_{200c}$ implicitly contains information about $M_{200c}$, which would also reduce the scatter.

The main implication of Table \ref{tab:PhotVSpec} is that, given an ideal training sample, the addition of galaxy line-of-sight velocities does not improve the accuracy over a purely photometric data set, i.e. one can obtain the same degree of scatter in the predicted mass with a photometric sample as a spectroscopic sample. \cite{Ntampaka2016} found a similar result using support distribution machines (SDM), where the scatter using the projected radial distribution of the galaxies was less than using the line-of-sight velocity distribution for their high-mass ($M>10^{15} \, \rm M_\odot$) cluster sample, and the same for their overall sample. One notable caveat to this statement is that our training sample does not contain any interlopers beyond $5r_{200c}$ of the cluster, due to the limited size of the high resolution volume in the simulation. Spectroscopic data may still be required to accurately remove more distant interloper galaxies, however \cite{Rozo2015} found when using redMaPPer that they could achieve cluster membership probabilities with photometric data that matched spectroscopic estimates at the per cent level.

\subsection{Inclusion of X-ray and SZ features}

\begin{figure}
	\centering
	\includegraphics[width=0.99\linewidth]{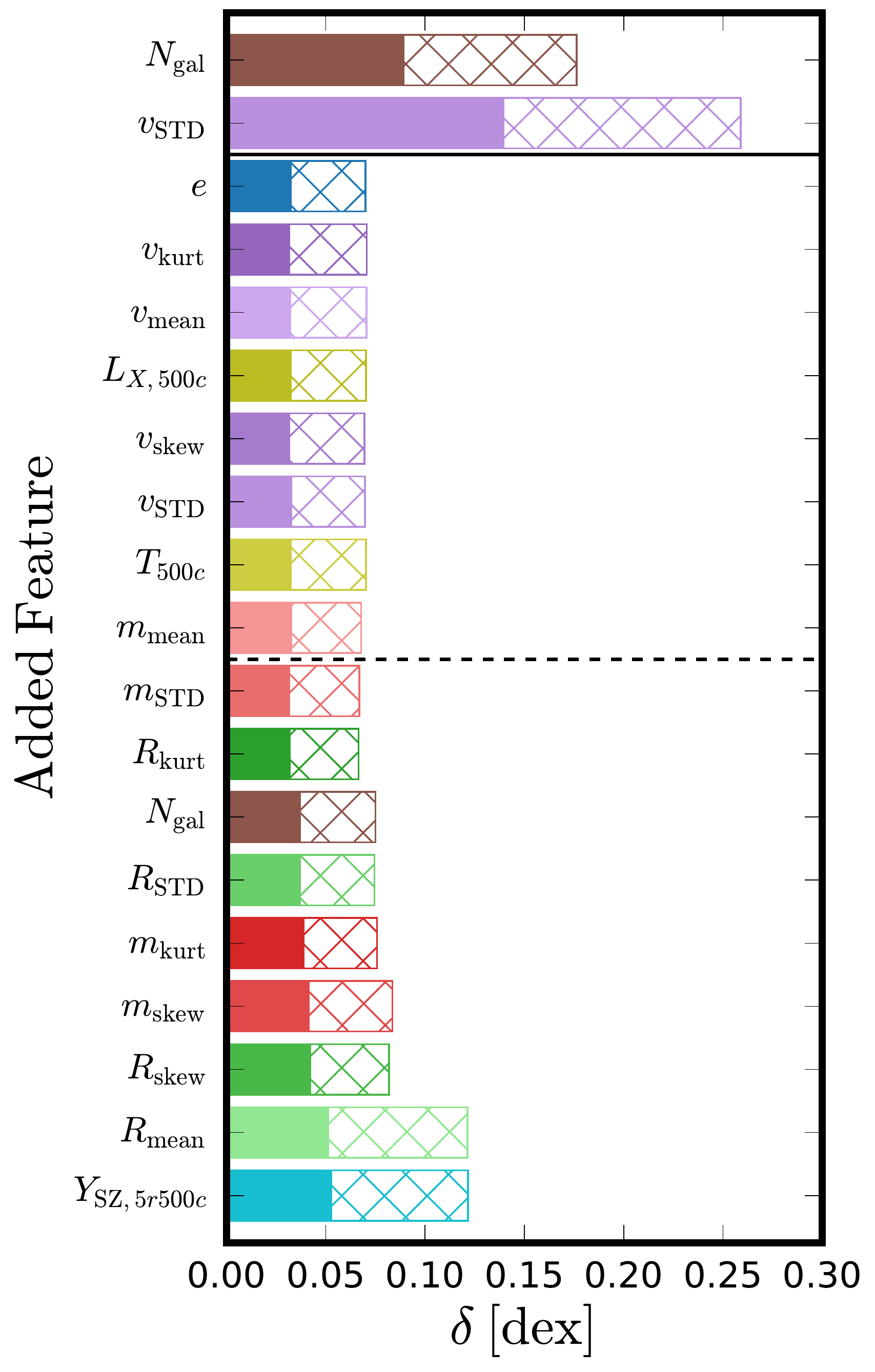}
	\caption{Comparison of the scatter, in dex, using different combinations of features. The ranges shown are the same as in Fig. \ref{fig:modelCompR200StelMass} and the colours match those in Fig. \ref{fig:FeatInd}. The order is determined by recursive feature elimination, where each row shows the scatter obtained from training a ridge regression model using all features below and at that point. The dashed horizontal black line shows the point where the minimum scatter (0.031 dex) is obtained; above the line the scatter is either the same or greater. Above the solid horizontal line, we show the scatter for two common observables: the number of galaxies in the cluster and their velocity dispersion, which were computed individually as a point of comparison. The cluster sample was selected within an aperture of $r_{200c}$.}
	\label{fig:FeatBuildUp}
\end{figure}

In addition to features derived from the galaxy population, one can include features from other observations to predict $M_{200c}$. We include the additional X-ray and SZ features in Figs. \ref{fig:FeatInd} and \ref{fig:FeatInd_Phys}, shown above the black dashed line. The $Y_{SZ}$ flux inside $5r_{500c}$ is the single most constraining feature, better than the galaxy-based features. The X-ray derived features, $L_{X, 500c}$ and $T_{500c}$, are taken within an observing aperture of $r_{500c}$, which will contain, implicitly, information about $M_{500c}$.

\begin{figure}
	\centering
	\includegraphics[width=0.99\linewidth]{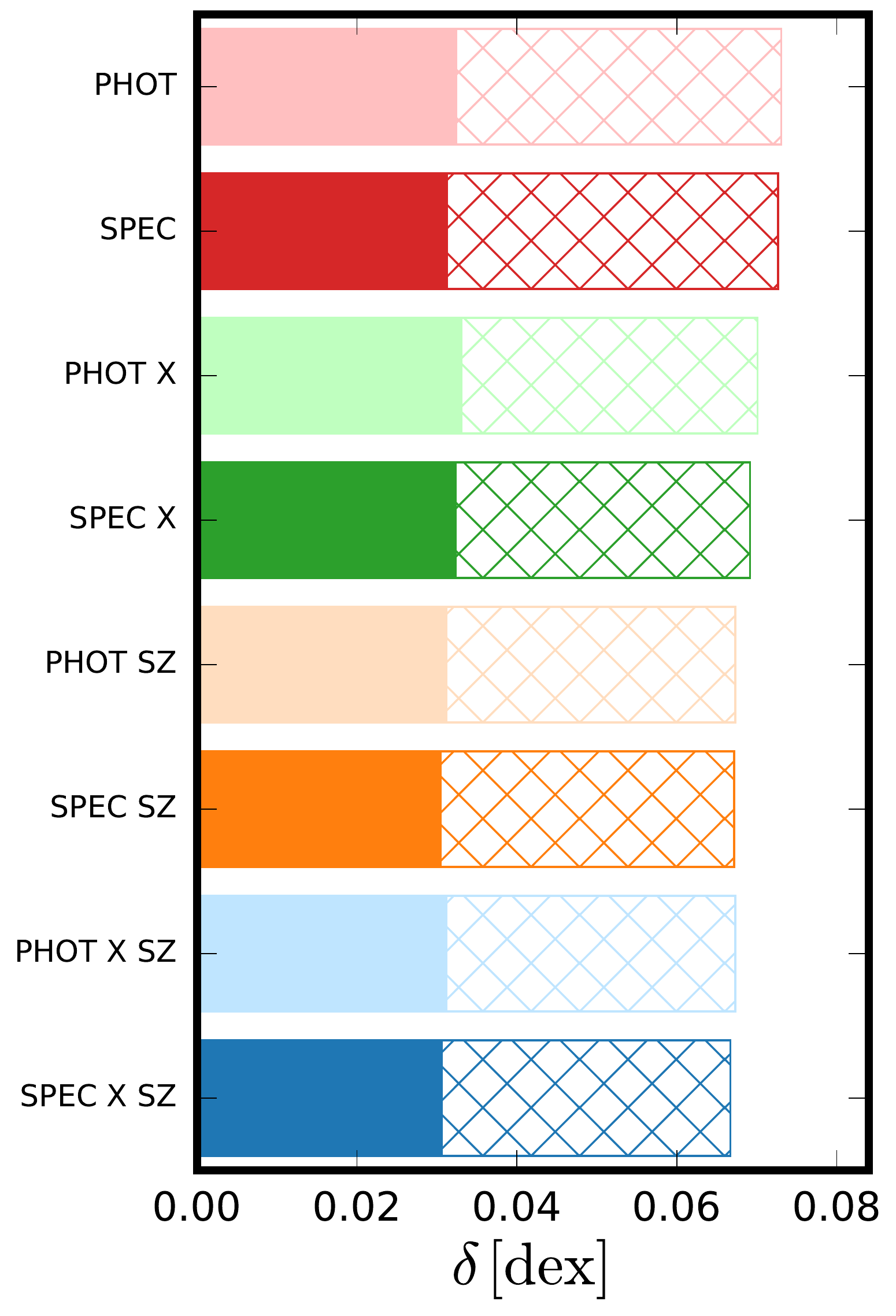}
	\caption{Comparison of the bias and scatter, in dex, using different feature sets, with galaxy derived features obtained within an aperture of $r_{200c}$. The ranges shown are the same as in Fig. \ref{fig:modelCompR200StelMass}. PHOT refers to a photometric dataset of galaxy positions and stellar mass; SPEC additionally includes line-of-sight velocity information; SZ marks the inclusion of the SZ flux within $5r_{500c}$; X denotes the inclusion of ICM derived properties, the bolometric luminosity and the surface temperature}
	\label{fig:FeatCompMtrue}
\end{figure}

Fig. \ref{fig:FeatBuildUp} shows the scatter when each feature is added in order of the amount of information contributed, according to RFE, using an aperture of $r_{200c}$. Starting from the bottom of the plot, each feature is added sequentially and the regression model is trained using it and all the features below. The black dashed line marks the point where there is no further improvement to the scatter (0.03 dex) through the addition of features. At the top of the plot, above the solid black line, we show the scatter for the velocity dispersion and number of galaxies when used individually (i.e. the same values as shown in Fig. \ref{fig:FeatInd}) for comparison. 

We stress that the exact order of features in Fig. \ref{fig:FeatBuildUp} is somewhat unstable due to the limitations of RFE as stated previously, particularly above the black dashed line. However, we can infer information from the general trends. As expected, the SZ flux and the projected radial distribution of galaxies both contribute a significant amount of information, as well as the higher order moments of the galaxy stellar mass distribution and the total number of galaxies present (richness). We can also see that the majority of the features are unnecessary; in the case where we use a fixed aperture, the number of rejected features decreases, though the majority only yield marginal improvements in the scatter.

As previously stated, there are some surprising results, particularly the galaxy velocity distribution not contributing much additional information. This leads to the question: given photometric observations of a galaxy cluster, how much is the scatter improved by including additional data/features? Fig. \ref{fig:FeatCompMtrue} shows how different combinations of observational feature sets affect the scatter in the predicted mass. SPEC and PHOT are as defined previously, SZ includes $Y_{SZ,5r_{500c}}$, while X includes $L_{b, 500c}$ and $T_{500c}$ as stated in Table \ref{tab:FeatBreakDown}. Each feature set is optimised using RFE before training on $M_{200c}$.

We see there is very little absolute difference between any combination of features. In general, the information loss from any singular feature is compensated for by other features contributing comparatively more unique information compared to when the removed feature was included. Discarding measures of the galaxy distribution does however lead to increased scatter. For example, removing the skew and kurtosis increases the scatter by 68 per cent, from 0.031 dex to 0.052 dex, with an error obtained via bootstrap resampling the data 1000 times of ${\sim}\pm 0.001$ dex for each.

\subsection{Training on mock X-ray and weak lensing masses} \label{ssec:ObsMasses}
\begin{table*}
	\centering
	\caption{The $1 \sigma$ percentile scatter, $\delta$, for different feature sets. Table \ref{tab:FeatBreakDown} summarises the quantities in each feature set. The uncertainties are the $1\sigma$ percentiles computed from 1000 bootstrap samples of the clusters. The left and right hand side of the table shows the scatter when taking galaxies within an aperture of $r_{200c}$ and 1.5 Mpc respectfully. The three columns in each are for the different target features, the true $M_{200c}$, $M_{200c}$ inferred from weak lensing, $M_{200c,\rm WL}$, and the hydrostatic mass inside $r_{500c}$ obtained from X-ray spectra $M_{500c,\rm X}$. For comparison we include the scatter obtained when finding the mass using the $\sigma-M$ relation, calibrated using the \textsc{MACSIS} clusters on the relevant target mass. }
	\label{tab:featAptScatt}
	\begin{tabular}{lccccccc}
		\hline
		& & $\delta_{r_{200c}}$ & & & $\delta_{1.5 \, \rm Mpc}$ & \\
		$\mathrm{Data \, set}$   & $M_{200c}$   & $M_{200c,\rm WL}$ & $M_{500c,\rm X}$ & $M_{200c}$   & $M_{200c,\rm WL}$ & $M_{500c,\rm X}$  \\
		\hline
		$\sigma - M$                 & $0.130\pm0.004$          & $0.117\pm0.004$                 & $0.182\pm0.006$                & $0.122\pm0.003$          & $0.117\pm0.004$                 & $0.177\pm0.006$                \\
		 PHOT                     & $0.032\pm0.001$          & $0.071\pm0.002$                 & $0.127\pm0.004$                & $0.065\pm0.002$          & $0.072\pm0.002$                 & $0.133\pm0.004$                \\
		SPEC                     & $0.031\pm0.001$          & $0.064\pm0.002$                 & $0.130\pm0.004$                & $0.064\pm0.002$          & $0.066\pm0.002$                 & $0.137\pm0.004$                \\
		PHOT X                   & $0.033\pm0.001$          & $0.067\pm0.002$                 & $0.088\pm0.003$                & $0.065\pm0.002$          & $0.065\pm0.002$                 & $0.090\pm0.003$                \\
		SPEC X                   & $0.032\pm0.001$          & $0.061\pm0.002$                 & $0.087\pm0.003$                & $0.064\pm0.002$          & $0.060\pm0.002$                 & $0.090\pm0.003$                \\
		PHOT SZ                  & $0.031\pm0.001$          & $0.069\pm0.003$                 & $0.123\pm0.005$                & $0.052\pm0.002$          & $0.070\pm0.002$                 & $0.131\pm0.005$                \\
		SPEC SZ                  & $0.031\pm0.001$          & $0.064\pm0.002$                 & $0.128\pm0.005$                & $0.051\pm0.002$          & $0.065\pm0.002$                 & $0.133\pm0.006$                \\
		PHOT X SZ                & $0.031\pm0.001$          & $0.063\pm0.002$                 & $0.086\pm0.003$                & $0.051\pm0.002$          & $0.066\pm0.002$                 & $0.089\pm0.004$                \\
		SPEC X SZ                & $0.031\pm0.001$          & $0.060\pm0.002$                 & $0.087\pm0.003$                & $0.052\pm0.001$          & $0.062\pm0.002$                 & $0.089\pm0.003$                \\
		\hline
	\end{tabular}
\end{table*}


A significant issue with the ML implementation as presented thus far is that it relies on knowing the true masses for a training set of clusters. This problem could be alleviated by training the model on a set of simulated clusters and then applying it to real data. Alternatively, one could train on a feature that is directly observable, such as the velocity dispersion \citep{Caldwell2016}, or on a set of clusters with a known mass, estimated via some other means. The first option could suffer from mis-calibrated subgrid models, cluster selection effects or instrumental limitations. At the very least for a new method, such as this, there needs to be some independent validation. In this spirit, we test the performance of our regression models when trained to reproduce masses obtained from mock weak lensing and X-ray observations. The driving philosophy behind this is that one could train on a high quality set of data and then apply that model to a larger, poorer quality, dataset, fulfilling a role similar to current scaling relations. While we have only trained to reproduce X-ray and weak lensing masses in this work, in principle one could train a model to reproduce any mass proxy, such as $L_\mathrm{X}$ or SZ signal.

Our results are summarised in Table \ref{tab:featAptScatt}, which shows the scatter for a variety of apertures, feature sets and target masses. Models are trained using either $M_{200c}$, the estimated value of $M_{500c}$ from X-ray analysis ($M_{500c,\rm X}$) or $M_{200c}$ from weak lensing ($M_{200c,\rm WL}$).
The $M_{500c,\rm X}$ masses are from the work of \cite{Barnes2017} and are computed by fitting the density and temperature profiles, parametrised by \cite{Vikhlinin2006}, and used to produce a hydrostatic mass profile to obtain $M_{500c,\rm X}$. The weak lensing masses are from \cite{Henson2017}, who fit the shear profile assuming an NFW mass distribution. Both $M_{500c,\rm X}$ and $M_{200c,\rm WL}$ are obtainable from observational data. This represents the scenario where one has, for example, a high quality X-ray sample of clusters, with overlapping optical data. One can then train a model using the smaller sample, before applying it to the larger optical catalogue.  In addition the models were trained only using galaxies within a projected radial distance of 1.5 Mpc from the cluster centre, removing any prior knowledge of the cluster size. 

We see a stark difference in the scatter depending on whether the model was trained using the true, X-ray or weak lensing mass. As we discussed above, training on the true value of $M_{200c}$ results in a uniform scatter of ${\sim} 0.03$ dex, irrespective of the feature set, within an aperture of $r_{200c}$. Training on $M_{200c,\rm WL}$ increases the scatter to ${\sim}0.65$ dex. We also see that the addition of spectroscopic data yields a $0.07$ dex improvement in the scatter over PHOT. Training on $M_{500c,\rm X}$ results in a significant increase in scatter, with the SPEC feature set resulting in a scatter of 0.13 dex, compared to 0.064 dex or 0.031 dex when training with $M_{200c, \rm WL}$ and $M_{200c}$ respectively. While it is not surprising that the scatters qualitatively increase, a jump of ${\sim}19$ per cent to ${\sim}35$ per cent is rather large. This indicates that, while in principle one could train using any mass estimate or cluster property, it must be chosen with care. Notably, fixing the aperture to 1.5 Mpc increases the scatter by only a small amount when training on $M_{500c, \rm X}$ or $M_{200c, \rm WL}$, but the scatter around the true mass nearly doubles in dex for many cases.  

Table \ref{tab:featR500Scatt} shows the scatter in dex when training to reproduce $M_{500c, \rm X}$ and $M_{500c}$ with a fixed aperture around each cluster of 1.5 Mpc. For reference the average true value of $r_{500c}$ for the \textsc{MACSIS} clusters is 1.23 Mpc, and 1.93 Mpc for $r_{200c}$. The only difference between the two columns in Table \ref{tab:featR500Scatt} is the target feature, removing the possibility that the increase in scatter with $M_{500c,X}$ is due to predicting a quantity based on $M_{500c}$ rather $M_{200c}$. We can see that the scatter improves significantly with the addition of more features. The PHOT feature set has a scatter of $0.072\pm 0.002$ dex compared to the $\rm SPEC\; X\; SZ$ set, $0.048 \pm 0.002$. Interestingly there is never a significant difference between PHOT and SPEC when training on the true value of $M_{500c}$. While there is an improvement from adding additional features to reproduce $M_{500c, \rm X}$ it is much less signifiant. The intrinsic scatter in $M_\mathrm{\rm 500c,X}$ appears to be the limiting factor in this case.

\begin{table}
	\centering
	\caption{The $1 \sigma$ percentile scatter, $\delta$, for different feature sets. This shows a subset of training masses, inside 1.5 Mpc. The formatting is the same as for Table \ref{tab:featAptScatt}.}
	\label{tab:featR500Scatt}
	\begin{tabular}{lcc}
		\hline
		&  $\delta_{1.5 \, \rm Mpc}$& $\delta_{1.5 \, \rm Mpc}$ \\
		$\mathrm{Data \, set}$   & $M_{500c,\rm X}$ & $M_{500c}$  \\
		\hline
		$\sigma - M$                 & $0.177\pm0.006$                & $0.130\pm0.004$          \\
		PHOT                     & $0.133\pm0.004$                & $0.072\pm0.002$          \\
		SPEC                     & $0.137\pm0.004$                & $0.071\pm0.002$          \\
		PHOT X                   & $0.090\pm0.003$                & $0.056\pm0.002$          \\
		SPEC X                   & $0.090\pm0.003$                & $0.053\pm0.001$          \\
		PHOT SZ                  & $0.131\pm0.005$                & $0.060\pm0.002$          \\
		SPEC SZ                  & $0.133\pm0.006$                & $0.058\pm0.002$          \\
		PHOT X SZ                & $0.089\pm0.003$                & $0.048\pm0.002$          \\
		SPEC X SZ                & $0.089\pm0.003$                & $0.048\pm0.002$          \\
		\hline
	\end{tabular}
\end{table}

\subsection{Dependence on baryonic physics}
 
Finally, we now consider the effects of baryonic physics on the model accuracy. We are interested in whether the inclusion of baryonic physics alters the scatter and how robust the ML models are to a change in subgrid physics (i.e. trained on DMO simulations and then applied to hydrodynamical clusters). The \textsc{MACSIS} clusters were simulated both with and without baryonic physics. To quantify the effect of baryonic physics on our feature set we train a ridge regression model using the \textsc{MACSIS-DMO} clusters and then apply that model to the \textsc{MACSIS-GAS} sample. We also computed `GAS to GAS' and `DMO to DMO' estimates for comparison. Throughout this section we use the the SPEC feature set exclusively, using subhaloes to represent the galaxies. In order to avoid any unrealistic correlations we train the model using 90 per cent of the DMO clusters and test on the remaining 10 per cent of clusters using the \textsc{MACSIS-GAS} sample. This way the DMO variant of a cluster will never be in the training sample when the \textsc{MACSIS-GAS} version is used to test the model. Similarly, we project the clusters along three orthogonal axes, but perform the training and testing analysis entirely separately for each projection. We only combine the projected samples once all predicted masses have been obtained.

\begin{figure}
	\centering
	\includegraphics[width=0.99\linewidth]{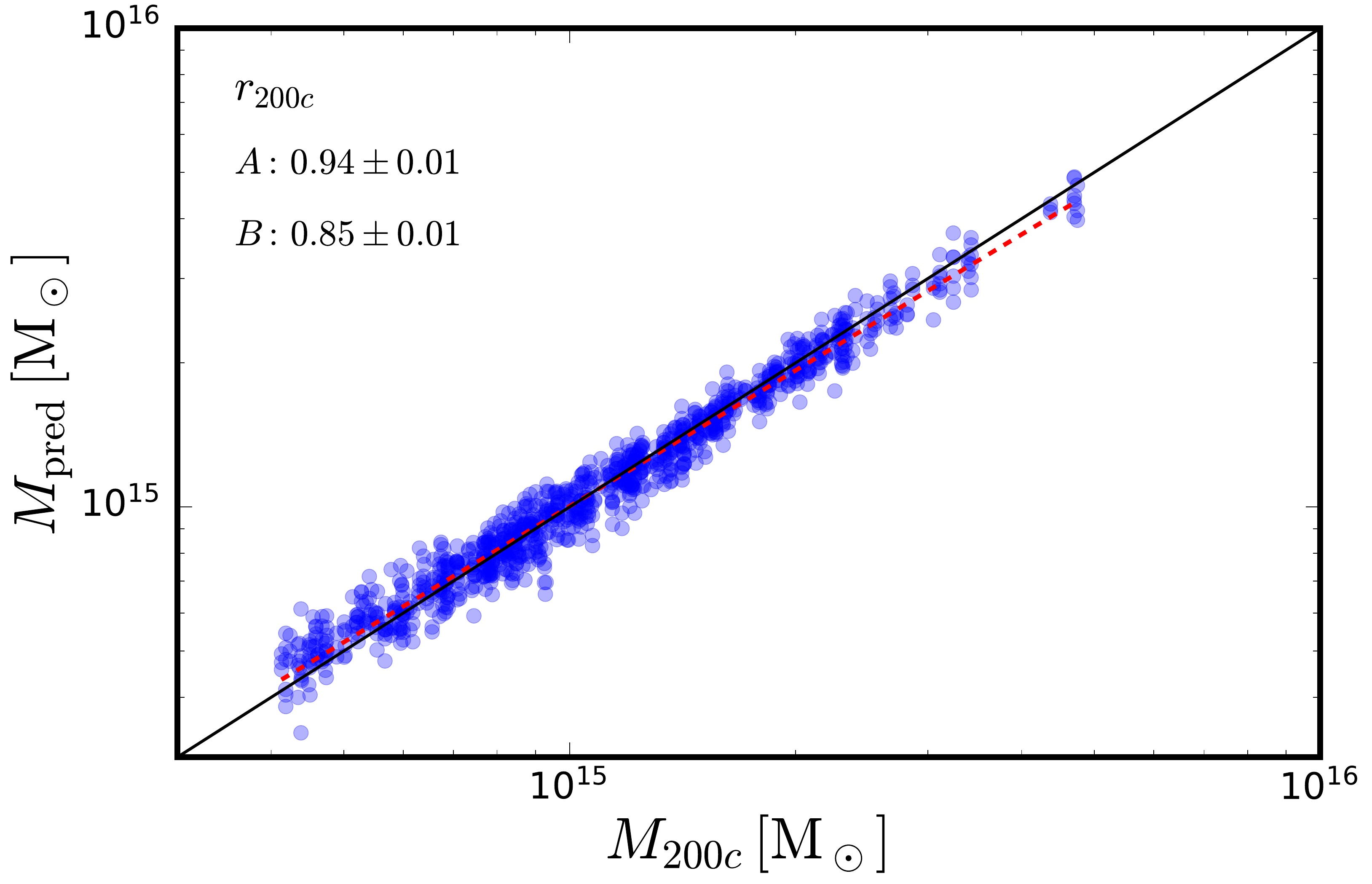}
	\includegraphics[width=0.99\linewidth]{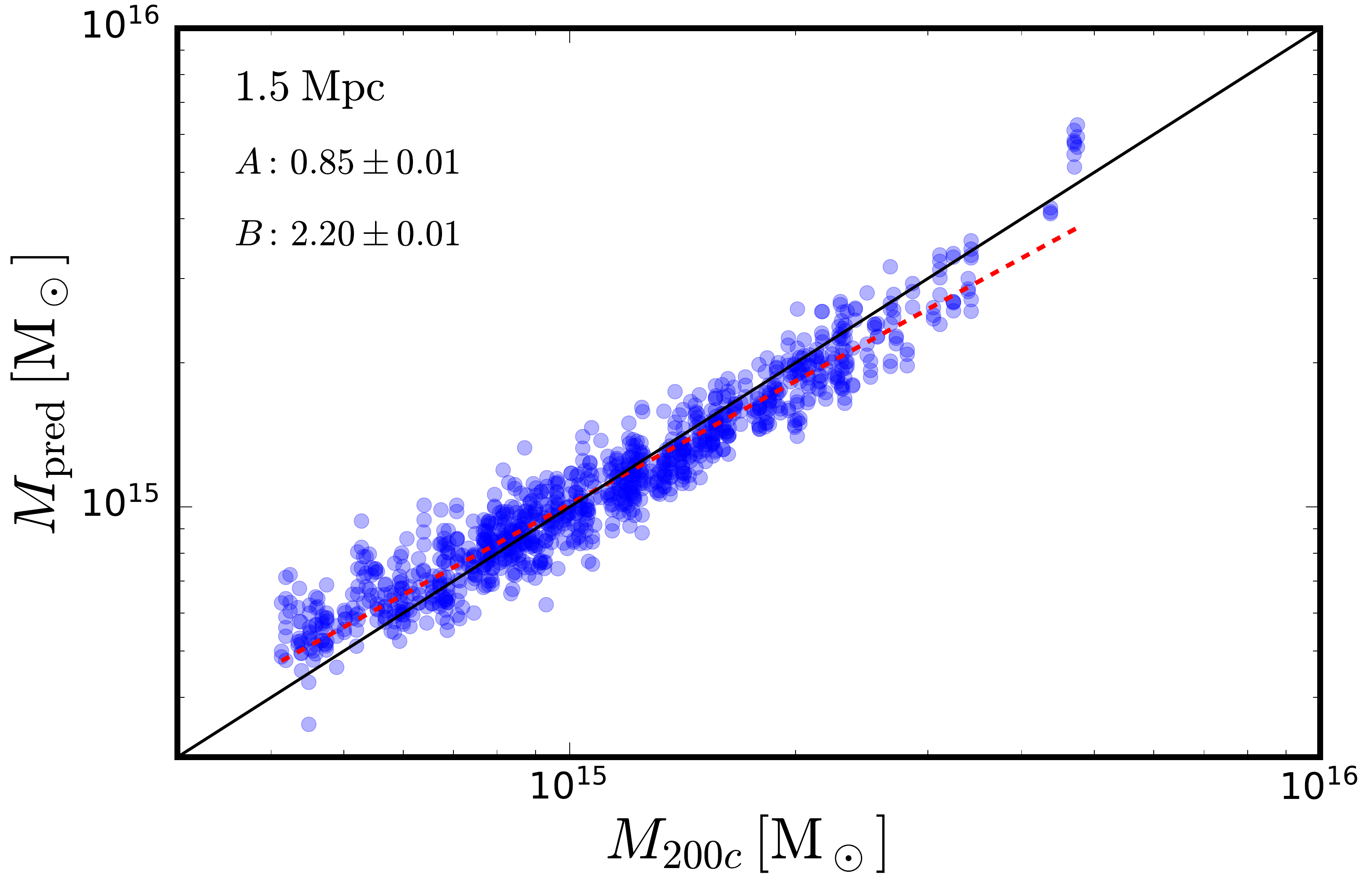}
	\caption{Top: Comparison of the predicted mass using a ridge regression model trained on DMO simulations of the \textsc{MACSIS} clusters, which is then tested using the hydrodynamical versions of the clusters. The dashed red line shows the best power law fit to the predicted masses, $\log \left( M_\mathrm{pred}/\mathrm{M_\odot} \right) = A \log \left( M_\mathrm{200c}/\mathrm{M_\odot} \right) + B$. Bottom: Same as above, though the aperture is fixed to 1.5 Mpc, rather than $r_{200c}$.}
	\label{fig:DMOComp}
\end{figure}

\begin{table}
	\centering
	\caption{The $1 \sigma$ percentile scatter, $\delta$, for when training and applying ML models to different simulations. The formatting is the same as for Table \ref{tab:featAptScatt}. The first part of the row label denotes whether the model was training using a DMO or GAS simulation, the latter half the type of simulation the model was applied too. The two columns are using galaxies within $r_{200c}$ and 1.5 Mpc respectively.}
	\label{tab:DMOvGAS}
	\begin{tabular}{ccc}
		\hline
		 Train-Test& $\delta_{r_{200c}}$  &   $\delta_{1.5 \, \rm Mpc}$ \\
		\hline
		GAS-GAS                  & $0.037\pm0.001$          & $0.055\pm0.002$          \\
		DMO-DMO                  & $0.037\pm0.001$          & $0.062\pm0.002$          \\
		DMO-GAS                  & $0.037\pm0.001$          & $0.064\pm0.002$          \\
		\hline
		
	\end{tabular}
\end{table}

Table \ref{tab:DMOvGAS} shows the scatter in dex for the six cases, note that the GAS-GAS values differ from those in Table \ref{tab:PhotVSpec} due to the cut in total rather than stellar mass. We applied a mass cut to the subhaloes of $5\times 10^{11} \; \mathrm{M_\odot}$, which approximately corresponds to a stellar mass cut of $10^{10} \; \mathrm{M_\odot}$. The scatter is the same as for a model trained on the \textsc{MACSIS-GAS} clusters for an aperture of $r_{200c}$, however, scatter for the 1.5 Mpc fixed aperture GAS-GAS case is smaller than for either DMO-DMO or DMO-GAS by ${\sim}0.007$ dex. This is a small difference corresponding to ${\sim} 1.5$ per cent, this is likely due to the MACSIS-GAS clusters containing 1.32 times more galaxies within 1.5 Mpc than MACSIS-DMO but only 1.15 times more at $r_{200c}$.

Fig. \ref{fig:DMOComp} shows the predicted $M_{200c}$ masses against the true values for clusters in the MACSIS-GAS runs, for the two apertures ($r_{200c}$ and 1.5 Mpc). The DMO training set recovers the true mass of the hydrodynamical clusters as accurately as for the DMO clusters, and introduces no significant bias (${\sim}1$ per cent). For the 1.5 Mpc aperture we see that the high mass clusters are underestimated.

We note a few important caveats: selecting lower total mass cuts introduces a bias into the recovered \textsc{MACSIS-GAS} mass estimates. This is primarily driven by the increased number of galaxies in the \textsc{MACSIS-GAS} sample, for a mass cut of $>10^{10} \, \mathrm{M_\odot}$ there are twice as many galaxies per cluster in the GAS sample. The difference is likely caused by the complex interplay of subgrid physics and limited resolution. We find the bias has converged by $M_\mathrm{cut}>5\times 10^{11} \, \mathrm{M_\odot}$, though it is difficult to validate as above this limit many clusters do not contain the minimum 10 galaxies. We therefore present this as an indication that DMO simulations can be used to train mass estimation models to be used in systems with baryonic physics, but more work is needed to validate this claim.



\section{Summary and Discussion} \label{conc}
In this work we have presented a novel way to use machine learning techniques to recover galaxy cluster masses. Features for each galaxy cluster are created from the observational properties of the clusters, such as the line-of-sight velocity dispersion, or the number of galaxies within an aperture. The features containing the most information are then identified via recursive feature elimination and this subsample of features is then used to train a ridge regression model. In this work we train the model on 90 per cent of the clusters, where the mass (or some proxy) of each cluster is known and then predict the mass of the remaining 10 per cent. When comparing the ML result to the $\sigma - M$ relation with a fixed aperture, we find the scatter in the predicted mass is reduced by a factor of 2, similar to that found in \cite{Ntampaka2016}, who make use of support distribution machines which use the distribution of the line-of-sight velocities and projected radial position of galaxies as features. Our main findings can be summarised as follows:

\vspace{-0.1cm}
\begin{enumerate}
	\item We find ridge regression to be the optimal regression algorithm, although the simple ordinary linear regression method performs similarly well when combined with recursive feature elimination (Figs. \ref{fig:RFE} \& \ref{fig:modelCompR200StelMass}).
	\item Compared to the $\sigma - M$ relation the scatter in the recovered mass is reduced from $0.130 \pm 0.004$ dex  (${\sim}35$ per cent) to $0.031 \pm 0.001$ dex (${\sim}7$ per cent), for clusters contaminated by interlopers (Table \ref{tab:PhotVSpec}).
	\item The addition of line-of-sight galaxy velocities does not improve the scatter in recovered mass when training on $M_{200c}$ (Table \ref{tab:PhotVSpec}).
	\item The skewness and kurtosis of the galaxy distributions contribute a significant amount of information and should be included for optimal results. The single most powerful feature is the SZ flux inside $5r_{500c}$ (Figs. \ref{fig:FeatInd} \& \ref{fig:FeatBuildUp}).
	\item We find no significant improvement over the PHOT feature set when including additional features from other observations, such as SZ or X-ray, when fitting for the true value of $M_{200c}$ (Fig. \ref{fig:FeatCompMtrue})
	\item Choosing to fit the estimated $M_{200c}$ mass from weak lensing or the hydrostatic $M_{500c}$ results in increased scatter of $0.063\pm 0.002$ dex and $0.130 \pm 0.004$ dex relative to fitting the true $M_{200c}$ mass, $0.031 \pm 0.001$. In general we find the hydrostatic $M_{500c}$ a difficult feature to reproduce and would recommend training a model using $M_{200c,\rm WL}$. This is likely to be due to the intrinsic scatter in the X-ray observations, not the change in aperture from $M_{200c}$ to $M_{500c}$ (Tables \ref{tab:featAptScatt} \& \ref{tab:featR500Scatt}).
	\item We find that training a model using DMO simulations and then estimating the masses of clusters run with hydrodynamics (GAS) gives the same scatter as estimating the masses of DMO clusters (Fig. \ref{fig:DMOComp} \& \ref{tab:DMOvGAS}). The scatter is slightly larger than for GAS-GAS training and testing, though the galaxy selection criteria differ. There is a bias introduced when estimating the mass of \textsc{MACSIS-GAS} clusters using a \textsc{MACSIS-DMO} trained model which is dependent on the mass cut used for the galaxies; this is likely due to the complex interplay between the subgrid model and the mass resolution.
	
\end{enumerate} 

The ML techniques that we have presented in this paper have several caveats that we shall explicitly state here. The first is that we present models that have been trained using the true value of $M_{200c}$. Clearly, in an observational setting one would never have access to the true mass of the cluster and so the models would need to be trained using either a mass proxy or some other mass estimate, such as the hydrostatic mass. We mimic such a scenario in Section \ref{ssec:ObsMasses}, with promising results when using mass estimates derived from weak lensing analysis. However, if the training mass is biased, then the predicted masses from the ML model will also share that bias. One could attempt to circumvent this by training a model using simulations and then applying that to observations directly, perhaps with a validation sample of clusters with masses estimated independently. This would require all of the observational systematics to be properly modelled and accounted for as well as the explicit assumption that the simulated clusters accurately represent reality.

We also note that while our sample of clusters include projection effects and interlopers, due to the nature of `zoom' simulations only relatively nearby ($<5 r_{200c}$) interlopers are included due to the limited simulation volume. Planned simulations in the near future will allow for the creation of light cones to properly mimic the inclusion of interlopers. 

The strength of the ML approach suggested in this work is that it can easily include multiple features at once to create a pseudo-scaling relation. This allows one to recover one feature, in this case the cluster mass, from a set of available features. This has the effect of drastically reducing the scatter compared to, say, the velocity dispersion mass relation.

In conclusion, ML algorithms, and in particular ridge regression, are a promising new approach to cluster mass estimation. The models can either be trained using simulated galaxy clusters, or by using observational data. To train a model using observational data one requires a good sample of galaxy clusters with estimated masses obtained via weak lensing for example. This would form the training set used to train the ML model, which would then be applied to a larger cluster catalogue which overlaps with the initial training data. Future work should focus on making more accurate mock observations to replicate the identification of clusters and member selection. This would likely cause a difference in the PHOT and SPEC feature sets once the entire observing pipeline is replicated. Identification of additional features, such as substructure and colour measurements could also improve the scatter.

\section*{Acknowledgements}
This work used the DiRAC Data Centric system at Durham University, operated by the Institute for Computational Cosmology on behalf of the STFC DiRAC HPC Facility (www.dirac.ac.uk). This equipment was funded by BIS National E-infrastructure capital grant ST/K00042X/1, STFC capital grants ST/H008519/1 and ST/K00087X/1, STFC DiRAC Operations grant ST/K003267/1 and Durham University. DiRAC is part of the National E-Infrastructure. DJB and STK acknowledge support from STFC through grant ST/L000768/1. TJA is supported by an STFC studentship. The authors thank Monique Henson for providing the weak lensing masses for the \textsc{MACSIS} clusters. This work benefited greatly from the availability of the \textsc{scikit-learn} python package \citep{sklearn}.


\bibliographystyle{mnras}
\bibliography{BibMain}





\bsp	
\label{lastpage}
\end{document}